%% file: 00_paper.tex
\documentclass[conference]{IEEEtran}
\IEEEoverridecommandlockouts
\usepackage{cite}
\usepackage{amsmath,amssymb,amsfonts}
\usepackage{algorithmic}
\usepackage{graphicx}
\usepackage{textcomp}
\usepackage{xcolor}
\usepackage{stfloats}
\usepackage[english]{babel}
\usepackage{svg}
\svgsetup{inkscapelatex=false}
\usepackage{url} 
\usepackage[hidelinks]{hyperref}
\def\BibTeX{{\rm B\kern-.05em{\sc i\kern-.025em b}\kern-.08em
    T\kern-.1667em\lower.7ex\hbox{E}\kern-.125emX}}
    
\newcommand{\qeng}[1]{\grqq#1\grqq}

\newboolean{review}

\newcommand{\reviewversion}[2]{
	\ifthenelse{\boolean{review}}
	{ #2 }
	{ #1 }
}

\setboolean{review}{false}

\begin{document}

\title{Blockchain Lifecycle Prediction - Dead Coins}

\reviewversion{
    \author{
        \IEEEauthorblockN{Uwe A. Kuehn} 
        \IEEEauthorblockA{\textit{Open Distributed Systems} \\
        \textit{Technical University of Berlin}\\
        Berlin, Germany \\
        0000-0002-1595-8401}
        \and
        \IEEEauthorblockN{Syed Muhammad Adnan} 
        \textit{Technical University of Berlin}\\
        Berlin, Germany \\
        m.syed@campus.tu-berlin.de}
}
{    
    \author{
        \IEEEauthorblockN{1\textsuperscript{st} Given Name Surname}
        \IEEEauthorblockA{\textit{dept. name of organization (of Aff.)} \\
        \textit{name of organization (of Aff.)}\\
        City, Country \\
        email address or ORCID}
    \and
        \IEEEauthorblockN{2\textsuperscript{nd} Given Name Surname}
        \IEEEauthorblockA{\textit{dept. name of organization (of Aff.)} \\
        \textit{name of organization (of Aff.)}\\
        City, Country \\
        email address or ORCID}
    }
}

\maketitle

\begin{abstract}
The cryptocurrency ecosystem has experienced extraordinary growth alongside an equally remarkable rate of failure, with over 52 percent of all tokens launched since 2021 ceasing to trade by early 2025. Despite the scale of this phenomenon, predictive modeling of cryptocurrency death remains an underdeveloped area of research, constrained by definitional ambiguity, data scarcity, and the absence of granular lifecycle frameworks. This work investigates whether the failure of cryptocurrency assets can be predicted using publicly available market data and deep learning methods.

A Long Short-Term Memory (LSTM) recurrent neural network was trained on 90-day sequences of daily reference price and estimated market capitalization for 82 cryptocurrency assets (41 alive and 41 dead), sourced from Coin Metrics over the period 2020 to 2026. The model was evaluated using a strictly chronological train-test split to prevent look-ahead bias. The LSTM classifier achieved in best cases a Receiver Operating Characteristic Area Under the Curve (ROC AUC) of 0.98 on the held-out test set. Finally, the model was applied to unseen data, and it was observed that the ROC AUC decreased between 0.59 and 0.65.

The findings demonstrate that temporal patterns in price and market capitalization alone contain sufficient discriminative signal to identify assets on a trajectory towards economic inactivity. Diagnostic analyzes reveal that failing assets exhibit gradual value erosion and elevated volatility in the months preceding inactivity, rather than sudden catastrophic collapse. This work also documents the practical infeasibility of a multi-stage lifecycle model under current data conditions and justifies the transition to a binary classification approach.

Limitations including the small number of dead coins in the sample, single-provider data source, and restricted feature set are discussed, alongside directions for future research encompassing feature expansion, horizon-specific labeling, and model interpretability.
\end{abstract}


\input{01_introduction}
\input{01_related_work}
\input{02_methodology}
\input{03_results}
\input{04_future_work_conclusion}

\bibliographystyle{abbrv}
\bibliography{references}
\end{document}

%% file: 01_introduction.tex
\section{Introduction}
Since the launch of Bitcoin in 2009~\cite{nakamoto2008}, the cryptocurrency ecosystem has experienced exponential growth. By April 2025 approximately 17,151 cryptocurrencies have existed, and though around 10,385 of them were deemed to be active~\cite{kumar2025, kumar2026}. This rapid expansion has attracted investors, developers, and speculators, but it has also created conditions conducive to the emergence of low-quality projects. These coins become inactive or are abandoned in large numbers and this leads to a phenomenon known by practitioners as dead coins. Dead coins may be characterised as the crypto values which no longer have a measurable trading volume, development or localization. The causes of this are very different: some projects are lost to rug pulls or even fraud, others fail due to the failure of the design and unsustainable crypto value economics, and many just lost interest or funding~\cite{fantazzini2022}. This results in very low survival rates for crypto assets.

The size of failures in the industry is pointed out by CoinGecko's statistics. Of nearly seven million tokens that had been listed on its GeckoTerminal platform by early 2025, 52.7 percent were no longer trading. 1.8 million tokens failed in the first quarter of 2025 alone, and almost 1.4 million projects died in 2024~\cite{coingecko53pct}. It is not limited to the micro cap experimentation: there have been high-profile collapses like TerraUSD (UST) and Luna in 2022 that erased billions of dollars~\cite{howmanydied}. This was not the first time. Previous studies found that approximately 59 percent of the evidence of work coins issued to date (before 2014) had defaulted by 2018~\cite{grobys2019}. The scale of failures increased substantially with the rise of token creation platforms such as \qeng{Pump.fun} in 2024, which significantly lowered the barrier to launching meme coins.

These statistics emphasise the need to detect coins that fail early enough. Dead coins are not just historical relics. They represent real financial losses for investors who may have committed significant capital. When the trading volumes are gone, so is liquidity, and exiting positions becomes practically impossible. An anticipatory framework that identifies at-risk tokens could assist investors and exchanges in managing exposure, help regulators monitor market health, and enable researchers to better understand the dynamics of cryptocurrency markets.

%% file: 01_related_work.tex
\section{Problem and Contribution}
One problem we observed is that there are no predictive models that determine crypto assets by different life cycle levels. Even though a number of studies approximate the likelihood that a cryptocurrency will default or go dead, they do not offer a model that defines a project's position on the failure curve. We developed a six-stage model to trace the development of phases that started with the inception and went through the growth and stagnation stage to the decline and terminal failure, and finally the ghost coin status. Unfortunately, no academic literature brings the stages together to demonstrate a consistent model. Rather, the literature talks of parts of the life cycle. Feder et al. and Schmitz and Hoffmann are examples of which criteria of death status are delineated~\cite{gandal2021}; Fantazzini and Zimin are examples of which credit risk models predict the probability of death based on the data about coin price and volume~\cite{sakinoglu2023}. All these works do not even make a distinction between the intermediate stages like stagnation or decline. As such, the gap in the research on stage-based diagnosis of crypto failures exists. Thus why we present a six-stage model to categorise the lifecycle stages, and we present a prediction model for detecting the transition from alive to dead crypto assets. Figure \ref{fig:lifecycle} illustrates our 6-stage cryptocurrency lifecycle model, and we will describe it in a separate section.

\section{Background and Related Work}
The introduction of Bitcoin in 2009 opened the gateway to a new generation of digital currencies whose quantity is increasing exponentially. According to Gatabazi et al. in an early study, the success of Bitcoin led to the introduction of other coins. In 2021, there are over 6,000 crypto assets, of which approximately 2,000 are already inactive or dead~\cite{gatabazi2022}. The authors note that the proliferation of new coins increased investment risk, and to a certain extent, this increase was propelled by the proliferation of coins, which was facilitated by initial coin offerings (ICOs), which forced some governments to prohibit cryptocurrency trading~\cite{gatabazi2022}. The digital asset ecosystem has been growing despite regulatory headwind. According to CoinGecko statistics, by 2025, almost 7 million crypto assets and tokens were registered in its GeckoTerminal service compared to 428,383 in 2021. During this time, 3.7 million projects ended trading, i.e.\ 52.7 of the tokens ended up failing. These failures increased in 2024 and 2025; CoinGecko records 1.8 million tokens failed in the first quarter of 2025 alone~\cite{coingecko53pct}.
Such statistics indicate the extremely high mortality of crypto assets and the importance of systematic lifecycle frameworks that describe the distinct stages a cryptocurrency asset passes through. According to survival analysis of cryptocurrencies, project attributes determine how long they will last. Gatabazi et al.\ applied non-parametric survival techniques to find that coins constructed on standalone blockchains (as opposed to tokens on existing chains such as Ethereum) had a higher risk of collapse. Tokens released during the 2013 and 2017 boom better survived than those released between 2009 and 2013. Projects with known headquarters had better survival chances~\cite{gatabazi2022}. These results suggest that background infrastructure, launch period and project openness have a significant impact on a coin's longevity.

\subsection{Crypto Assets and their Categories}
By analysing crypto assets, we have to mention that not every crypto asset is the same as another crypto asset, and often synonyms have been used in a wrong way, so that there can be a misunderstanding of what is really meant. For some reason, digital asset and crypto value are synonyms for crypto asset, and are used as an umbrella term to mean just all of them~\cite{EGLOFF2026105}.
In this work, we want to use a simple and small definition of crypto assets. Cryptocurrencies are those technologies which build native coins based on DLT, like Bitcoin, or Ethereum. These are historical and fundamental blockchain networks built on distributed ledger technologies (DLT). The trade with assets is very risky, since the price for 1 native coin is unstable. In contrast to this, there are crypto assets called \qeng{stablecoins}, which, as the name promises, are regarding the price for 1 stable coin. Usually references stable coins to fiat currencies like US-DOLLAR or EURO if asset-backed.  Tokens are different to the native assets and stablecoins, as they usually require an existing DLT and exist on those networks only. There are a lot of different tokens available, listed here~\cite{cryptobook2026}, and not discussed further.

\subsection{6-staged-lifecycle}
\begin{figure*}[!ht]
    \centering        
    \includegraphics[width=\textwidth]{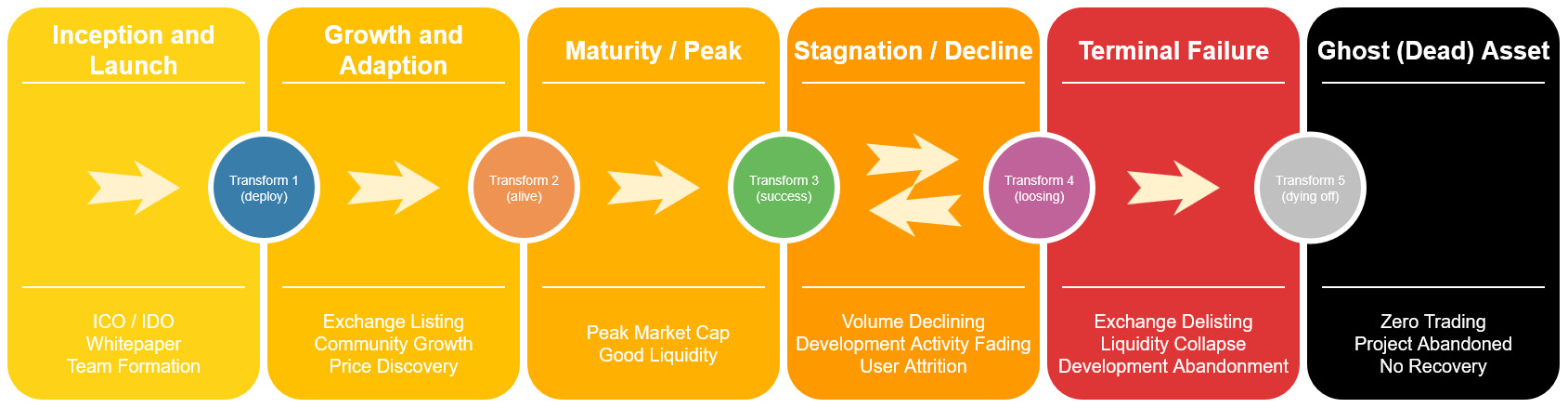}
    \caption{6-stage crypto asset lifecycle}
  \label{fig:lifecycle}
\end{figure*}

The lifecycle of a cryptocurrency is initiated much earlier than the trade of the assets. The guides to the industry have segmented the asset lifecycle in various stages: ideation and team formation, where the vision and market fit are formulated; whitepaper and tokenomics design; token sales (seed rounds, private sale, presale and public Initial Coin Offering (ICO) and Initial DEX offering (IDO); token generation and exchange listing; and post listing development and ecosystem growth~\cite{tseng2021}. The risks and rewards associated with each of the stages are different. The first phases have the highest upside yet the highest uncertainty and the second phases are more adoption and liquidity with not as much volatility. An important indicator to investors during the pre-launch phase is whitepaper quality, team credibility, and tokenomics fairness~\cite{cryptolifecycle}. This first step and the first transformation will be hard to predict, since usually no information can be received very soon. 
In addition to fundraising, the market cycles dictate the performance of tokens. Analysts identify four stages in a crypto market cycle---accumulation, markup, distribution and markdown. In the accumulation stage, the trade volume is average, the mood is negative and slowly moves towards better and a group of investors silently accumulates stock. Although these phases define the state of the market, they are similar to the life cycle of projects; assets tend to be launched during or immediately after bull runs and fail once the liquidity is cut off~\cite{alzahrani2019}. These cycle stages are connected by the transformation steps 2, 3 and 4 that we present in our lifecycle model in Figure~\ref{fig:lifecycle}. Finally, the crypto assets may fail and die. It is not very clear if there is a path back via the transform step 4, but once the coin is dead, and trading the assets have been stopped, nobody trusts this asset. That's why our model does not support reincarnation, but if a dead coin comes back, then it is covered by the transformstep 1.

\subsection{Dead coin predictive models}
What is the reason behind the failure of so many cryptocurrencies? Several factors contribute. Common causes include community disengagement, fraud, sustained low trading volume, and the disappearance of the project's online presence~\cite{deadcoinsnd}. According to Deadcoin repositories, most projects are either joke coins or rug pulls, launched without the intention of being used~\cite {deadcoinsnd}. According to a table drawn by CoinMarketCap, coins are considered dead when an extended period of time passes without a trading volume of at least 1 percent of that volume historically, or when the coin is not traded at all in 30 days or longer~\cite{deadcoins2024}. Crackdowns by the regulatory authorities, like the ban of ICOs by China, can lead to collapse too. Unhealthy tokenomics like oversupply, unrealistic returns or unfair distribution to insiders suppress investor confidence. Misaligned incentives in the form of technological vulnerabilities (hacks, bugs) and/or failure to achieve the original vision through decentralised forms of governance also contribute to failure~\cite{deadcoins2024}. Macro economic variables such as bear markets and interest rates increase, discouraging risk assets and hasten the death of marginal projects. Lastly, tokens can become outdated because of competition in saturated niches (e.g.\ DeFi, NFTs) and the appearance of better platforms.
Cryptocurrency that has become unprofitable is referred to as a dead coin. Nonetheless, it does not have a standard definition. According to Fantazzini~\cite{fantazzini2022}, the criteria proposed by professional websites and scholarly research may be different: some of them consider a coin to be dead when its price is less than a cent~\cite{fantazzini2022}, and others insist on an extremely low level of trading, the disappearance of nodes, loss of community and removal by exchanges. The first scholarly definition is the one given by Feder et al.\ (2018). They refer to the peaks of a history of coin prices when the seven-day rolling average trading is larger than the historical ones; a coin is abandoned when the average daily trading volume of a month is lower than 1 percent of the historical maximum. Schmitz and Hoffmann (2020) made it easier by calling a coin dead when its monthly trading volume declines below 1 of its historical peak and resurrected when it increases to above 10~\cite{fantazzini2022}.
Financial time series are known to be volatile and nonlinear. Most conventional statistical models (autoregressive integrated moving average (ARIMA) and generalised auto-regressive conditional heteroscedastic (GARCH)) that have been widely applied in market forecasting are essentially linear. According to the hybrid CNN-LSTM research article by Badar et al., such classical models do not easily capture long-range complex dependencies that define the price variation of Bitcoin~\cite{badar2025}. This is because they are unable to learn nonlinear patterns and hence they have poor predictive performance, especially when volatility is extreme. As a result, scholars have resorted to machine learning (ML) and deep learning (DL) algorithms to achieve greater accuracy~\cite{badar2025,chaim2019}.
Initial ML solutions to predict cryptocurrencies were based on such models as support vector machines (SVMs) and random forests. These models are an enhancement of the linear models, yet have difficulties with the sequential dependence in the time series data. With the advancement in research, recurrent neural networks (RNNs), particularly long short-term memory (LSTM) networks, gained popularity due to the use of memory cells that capture long-term dependencies~\cite{sebastiao2021} and deal with vanishing/exploding gradient problems. The hybrid CNN-LSTM article notes that LSTMs are specifically designed to obtain sequential correlations, and as a result, are highly suitable for predicting the prices of Bitcoin. Convolutional neural networks (CNNs) are used with LSTMs to extract spatial features of time series inputs, which leads to researchers creating hybrid CNN-LSTM models that combine spatial and temporal feature extraction. Badar et al.~\cite{badar2025} combined variational autoencoders (VAE) to learn the feature and Shapley additive explanations (SHAP) to understand the feature. Their model had a mean squared error (MSE) of 0.0002, a an R-squared mean absolute error (MAE) of 0.008 and a R-squared of 0.99, which was better than single CNN or LSTM models. The authors put forth that interpretability offered by SHAP can help in comprehending the importance of features and increases trust. These findings indicate that hybrid deep learning models have the ability to represent sophisticated dynamics in crypto markets and at the same time ensure transparency.
To make the architectural comparison concrete, Table~\ref{tab:arch_comparison} summarises the key structural distinctions between LSTM, BiLSTM, GRU, and CNN-LSTM models as they apply to the cryptocurrency survival prediction context. The critical dimension for this work is not peak accuracy reported in any single study, but suitability under the deployment constraints that define the problem: real-time forward-only inference, a fixed 90-day input window, a binary output, and a very small pool of confirmed dead assets. In Table~\ref{tab:arch_comparison} we compare used methods and list the ability of Reads Sequence Direction (RSD), Spatial Feature Extraction (SFE), Real-Time Forward Inference (RT FI), with our solution.
The statement by Konuk and Guvenir~\cite{konuk2025} that the LSTM showed lower and more variable performance than GRU in their study requires clarification, because it is superficially at odds with the decision to adopt LSTM in this work. Their finding was specific to a variable-length input setting (10 to 180 days of history) where GRU's simpler gating mechanism generalized more consistently. In the fixed-window setting used here (exactly 90 days for every window), this variability advantage of GRU disappears. Moreover, GRU reduces the temporal memory depth relative to LSTM by collapsing the cell state and hidden state into a single vector, which may sacrifice long-range dependency capture. Given that the pre-death signatures pan 180 days and show gradual multi-month patterns, the deeper memory architecture of LSTM is better matched to the temporal scale of the target phenomenon~\cite{cho2014,chung2014}.

\begin{table}
\centering
\caption{Comparative summary of architectures}
\label{tab:arch_comparison}
\footnotesize
\begin{tabular}{|p{1.7cm}||c|c|c|c|}
\hline
\textbf{Architecture} & \textbf{RSD} & \textbf{SFE} & \textbf{RT FI} & \textbf{Our work} \\ \hline
LSTM & fw only & No & Yes & Yes \\ \hline
BiLSTM & fw + bw & No & No  & No \\ \hline
GRU & fw only & No & Yes & No \\ \hline
CNN-LSTM & fw only & Yes &  Yes & No\\ 
& \footnotesize{(LSTM stage)} & \footnotesize{(CNN stage)} & &\\
\hline
\end{tabular}
\end{table}

%% file: 02_methodology.tex
\section{Methodoloy}
This work presents cryptocurrency survival prediction as a supervised binary prediction problem on multivariate daily market time series. For each cryptoasset $a$, the model ingests a fixed-length lookback window of the most recent $T = 90$ days of observations and outputs a probability score $p(y_a = 1 \mid X)$, where $y_a = 1$ denotes the \qeng{dead} class under the operational definition.

The interpretation of the score is that it is a rolling risk indicator: as new data are received on a daily basis the most recent 90-day window is recast and re-scored. The design is pragmatically designed to the dead-coin environment where indicators non-market (e.g.\ developer activity, governance metrics, or community signals) are usually inaccessible or no longer provided in the case of assets that end up failing. The model, therefore, depends on variables of the market that are always evident in assets.

The reason why a Long Short-Term Memory (LSTM) recurrent neural network is utilized is because it can train non-linear temporal effects in sequential data and is not as susceptible to vanishing gradients as vanilla recurrent neural networks as a result of its gated memory component~\cite{hochreiter1997}. The methodology selection in the chapter is motivated by the fact that 
\begin{enumerate}
    \item[(i)] it is necessary to avoid the look-ahead bias in the evaluation,
    \item[(ii)] it is necessary to minimize the survivorship bias in the feature availability, and
    \item[(iii)] it is necessary to obtain results that can be reproduced with the help of an understood labeling rule and a generally available market data.
\end{enumerate}

\subsection{Data source and Data collection}
Coin Metrics provided the API v4 on the official Python client to gather market data on a daily basis. Coin Metrics offers standard asset identifiers and curated market data such as reference prices and estimated market capitalisation metrics, to be comparable across exchanges/trading venues~\cite{coinmetricsnd}.

The range of the extraction window selected for prototype 2020-01-01 to 2025-11-30 at a daily frequency. In the second part of our analysis, we fetched data until 2026-05-30. Two market measures were requested on each asset
\begin{enumerate}
    \item[(i)] ReferenceRateUSD (reference price in USD) and 
    \item[(ii)] CapMrktEstUSD (estimated market capitalisation in USD).
\end{enumerate}
ReferenceRateUSD is built on the reference-rate structure of Coin Metrics that is supposed to be less vulnerable to outliers and manipulation than a single-exchange quote, whereas CapMrktEstUSD adheres to the industry norm of circulating supply times price~\cite{coinmetricsnd}.

The two variables of interest in the study are price and market capitalisation since they are freely available on both established and marginal assets and can be tracked in close real time by the exchanges, analysts, and investors. Although other predictors (e.g., trading volume, liquidity, on-chain activity, off-chain signals e.g., social media or GitHub activity) can be more effective, they are frequently unavailable, or inadequately defined, and not available on assets that later fall out of use. Limiting the feature set thus helps in replicability and also minimises feature-based survivorship bias.

The raw time series has missing values as a result of non-provider coverage, lack of reporting exchange, and thin trading. In each asset, the vacant values in every feature were filled forward and backward in such a way that the short gaps would not separate 90-day sequences. Following imputation, any missing rows containing feature values were eliminated in order to make all the windows where modeling was developed to be complete.

Forward/backward fill implicitly supposes that short jumps imply reporting noise and not actual economic jumps. To minimise the influence of this assumption, imputation is done within-asset (not across assets), and windows that remain with missing values after doing this are eliminated.

\begin{figure}
    \centering
    \includegraphics[width=0.45\textwidth]{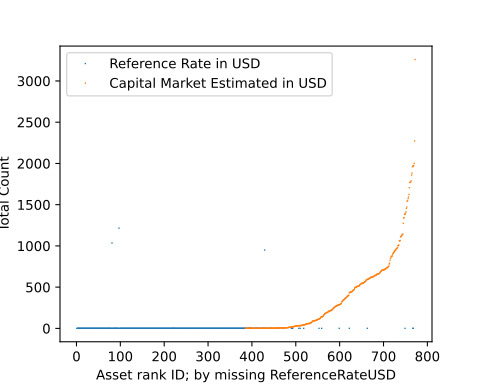}
    \caption{Absolute missing values}
  \label{fig:abs_missing_values}
\end{figure}

\begin{figure}
    \centering
    \includegraphics[width=0.45\textwidth]{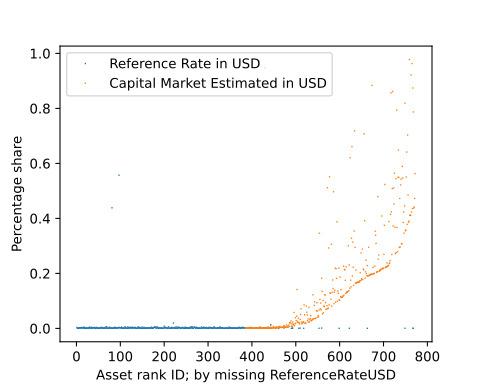}
    \caption{Relative missing values}
  \label{fig:rel_missing_values}
\end{figure}

A quantitative audit of the raw missingness, shown in Figure~\ref{fig:abs_missing_values} and~\ref{fig:rel_missing_values}, reveals that reference price (ReferenceRateUSD) had in total 1,000 missing daily observations, representing 0.63 percent of the total 157,860 rows, while estimated market capitalisation (CapMrktEstUSD) had 12,256 missing values, or 7.76 percent of the total. The substantially higher rate of missing market-cap observations reflects a known characteristic of the Coin Metrics API coverage: market capitalisation data is typically backfilled only once circulating supply estimates become available, and may lag or be absent for newly listed or thinly traded assets. Missing price values, by contrast, predominantly arise from non-trading days on exchange venues or short coverage gaps.

Forward fill propagates the last known valid observation forward in time, and backward fill covers any remaining gaps at the start of an asset's history. Both operations are performed exclusively within each asset's own time series, preventing any cross-contamination of values between different coins. After within-asset imputation, all rows that still contain at least one missing feature value are removed before sequence construction. This deletion step is justified because LSTM sequence learning requires complete input matrices: any remaining null values would cause silent numerical errors or require zero-imputation, which would misrepresent the true distributional properties of the input. The joint imputation-and-deletion strategy ensures that all 90-day windows presented to the model are fully observed, while retaining the maximum amount of data from assets with short but complete gaps.

As a robustness note, the aggregate row-level missingness rate of approximately 4.3 percent (combining both features) is modest relative to the dataset size, and the within-asset gap structure means that imputation errors are localised. Assets with structurally high missingness, such as coins that ceased reporting before the study period, effectively self-select out of the cohort through the 365-day inactivity labelling rule and the minimum-sequence-length filter applied during cohort construction.

While Figure~\ref{fig:abs_missing_values} shows the absolute amount of missing data, Figure~\ref{fig:rel_missing_values} shows the relative missing data, based on the total amount of fetched data for each coin. On the x-axis, we list the total available coins ordered by their CapMrktEstUSD. From that collection of data, we pick a maximum of 100 coins from the top 100 for alive and dead coins. On the y-axis, we show the total number of missing data points in Figure~\ref{fig:abs_missing_values}, and relative in Figure~\ref{fig:rel_missing_values}. The error we get from skipping missing data is for each coin less than 1 percent.

No single universal definition of a dead cryptocurrency exists. In prior literature and practice, operational definitions have traditionally been based on prolonged inactivity, delisting, or the loss of stable market quotes~\cite{fantazzini2022,ammann2022}. We mark an asset as dead when it does not have any Coin Metrics observations in the last 365-day period of the study period. That is, the final possible observation should not be more than a year before the end date of the dataset.
Thus, we labelled the data sets as dead (1) and alive (0), as illustrated in Formula~\ref{eq:dead_label}.
\begin{equation}
  y_a = 
  \begin{cases}
    1 & \text{if } t_{\text{last}}(a) \leq T_{\text{end}} - \Delta \\
    0 & \text{otherwise}
  \end{cases}
  \quad \text{where } \Delta = 365 \text{ days.}
  \label{eq:dead_label}
\end{equation}

Coin-level statistics of all assets (first date, last date, maximum market capitalization, and most recent market capitalization) were calculated. Two cohorts were then formed to create the modelling dataset
\begin{enumerate}
    \item[(i)] up to 100 dead assets (by rank) with the highest historical CapMrktEstUSD and 
    \item[(ii)] up to 100 live assets (by rank) with their most recent CapMrktEstUSD. 
\end{enumerate}
The design focuses on high-impact assets in each class and minimizes excessive class imbalance on the coin level and places the highest priority on assets that are the most relevant to exchanges and investors.
Practically, it is the fact that the constraint of up to 100 is important due to the fact that not every candidate dead coin has a history long enough to create stable 90-day sequences. The inactivity criterion, together with minimum-history, reduced the selected dataset to 121 assets (100 alive and 21 dead) in our prototype. While fetching data until 2026-05-30 we were able to label 41 dead coins.
In order to prevent look-ahead bias and leakage in time-dependent prediction tasks, model evaluation was done with a chronological split that ensures that all training samples come before the time period~\cite{hyndman2021,kapoor2023}. This configuration resembles a deployment situation where a model is trained with past historical data and applied to the future days that were not present during the training.
The end timestamp of windows before 2024-01-01 was used to test and put in the training pool, and those with an end timestamp on or after 2024-01-01 were put into the test set. Out of the training pool, an 80/20 stratified division was carried out to form a validation set to be used in early stopping and threshold tuning. Even though the validation split falls inside the training period, the test set is entirely out-of-time and gives the overall estimate of the generalisation performance. To have separation for test and train data in a real 80/20 (and others) distribution instead of fixed date splitting, we also did experiments with different split methods in the data until 2026-05-30. We calculated for each crypto asset in the training data the optimal date to reach the optimal distribution.

Both features were normalised by z-score and were only fitted on the training set and finally applied to the test and validation sets~\cite{bishop2006,pedregosa2011}:
\begin{equation}
  z = \frac{x - \mu}{\sigma}
  \label{eq:zscore}
\end{equation}
In this case, $\mu$ and $\sigma$ are the mean and standard deviation calculated on the basis of the training data. Scaler was adjusted by flattening time-varying training windows so that $\mu$ and $\sigma$ indicate the training distribution of every feature. Standardisation puts the price and market capitalisation on similar scales and stabilises the optimisation of neural networks.

Sliding windows may cause the unequal number of samples per class due to the alive assets occupying the entire observation window and dead assets being cut off. In order to minimise the bias on the majority class, cost-sensitive learning was adopted by weighting the classes during training~\cite{elkan2001,he2009}. The weighting heuristic that is balanced is:

\begin{equation}
  w_c = \frac{N}{K \cdot n_c}
  \label{eq:classweight}
\end{equation}

In this case, $w_c$ means the weight of class $c$, $N$ is the number of training samples, $K$ is the number of classes ($K = 2$), and $n_c$ is the number of training samples in a specific class~\cite{pedregosa2011}. The weights of classes were calculated based on the training split alone.
An LSTM network maps each $90 \times 2$ input sequence to a single probability. The LSTM maintains a memory state $c_t$ and uses input, forget, and output gates to control information flow~\cite{hochreiter1997}. The implemented architecture is: \texttt{LSTM(64, return\_sequences=True)} $\to$ \texttt{Dropout(0.2)} $\to$ \texttt{LSTM(32)} $\to$ \texttt{Dropout(0.2)} $\to$ \texttt{Dense(32, ReLU)} $\to$ \texttt{Dense(1, Sigmoid)}
Dropout regularisation reduces overfitting by randomly dropping units during training~\cite{srivastava2014}. The first LSTM layer returns the full sequence of hidden states, allowing the second LSTM layer to aggregate longer-range information into a compact representation. The final sigmoid unit outputs a probability in $[0, 1]$. Probabilistic outputs require a decision threshold $\tau$ to produce hard labels. Rather than fixing $\tau = 0.5$, $\tau$ was selected to maximise validation F1-score, which summarises the precision--recall trade-off via a harmonic mean~\cite{vanrijsbergen1979,rennie2004}. Thresholds in $[0.05, 0.95]$ with a step 0.05 were scanned on the validation set, and the best $\tau$ was then applied to the held-out test set. 
The experiments were coded in Python with NumPy and pandas to process and manipulate data, scikit-learn to preprocess and evaluate metrics~\cite {pedregosa2011,scikitlearn}, TensorFlow/Keras to train neural networks~\cite{abadi2015},~\cite{abadi2016} and Coin Metrics API client to access data. Random seeds were cemented (NumPy and TensorFlow) so as to enhance reproducibility. Additionally, we publish our selected results in a public GitLab~\cite{git_dead_coins_data} repository.

%% file: 03_results.tex
\section{Results}
In our experiments, we started with a prototype, which we have already described. In the table below, we list some interesting examples where we modified some details for data selection, training and testing the learned model.
\subsection{Experiments Overview}
In Table~\ref{tab:avdata} we show the setup of our different analyses. On the one hand, we modified the ratio of training vs. test data. The fix-date is set to \qeng{2024-01-01}. It is important to note that data points in the training data are only available before this date, and test data are only used beginning on this date.
A result of this decision is that some training or test data have fewer data points than we require for training or predicting the state of a crypto asset. We illustrated this behavior for 7 of 21 dead coins used in setup 0602 in Table~\ref{tab:av_data_fixed_split}. 
For the purpose of subsequent analyses, the decision was taken to transition from the fixed-date split to the more commonly employed percentage split. The presentation of a limited number of samples in Table~\ref{tab:av_data_optimal_split} for setup 0614 has been deemed sufficient. We calculated an optimal split date, and tested different ratios for train and test data to learn the impact on our analysis. Lastly, we extended the whole time span for data aggregation. We doubled the training data size for dead coins. The span \qeng{A} covers the data fetched until 2025-11-30, and the span \qeng{B} extends available data until 2026-05-30. Consequently, a greater number of dead coins were retrieved for the purpose of training.
\begin{table}[h]
    \centering
    \caption{Data used for training and test}
    \begin{tabular}{|c||c|c|c|c|c|c|c|}  \hline
       Setup Id & Span & Ratio & \# Alive & top & \# Dead & tres\\ \hline \hline
       0602 & A & fix-date & 749 & 100 & 21 & 0.60 \\ \hline
       0603 & A & 80:20 & 749 & 100 & 21 & 0.55 \\ \hline
       0604 & A & 90:10 & 749 & 100 & 21 & 0.30 \\ \hline
       0605 & A & 70:30 & 749 & 100 & 21 & 0.55 \\ \hline
       0610 & B & 80:20 & 740 & 41 & 41 & 0.50 \\ \hline
       0614 & B & 90:10 & 740 & 41 & 41 & 0.85 \\ \hline
    \end{tabular}
    \label{tab:avdata}
\end{table}

\begin{table}[h]
    \centering
    \caption{Days distance using fixed date splitting setup 0602}
    \label{tab:av_data_fixed_split}
    \begin{tabular}{|l||l|l|r|r|}  \hline
        Asset  & First Seen & Last Seen  & Before Split & After Split \\ \hline   \hline
        aion   & 2020-01-01 & 2023-03-04 & 1461  & -303  \\ \hline
        ant    & 2020-01-01 & 2024-11-08 & 1461  & 312   \\ \hline
        crpt   & 2020-01-01 & 2024-09-19 & 1461  & 262   \\ \hline
        cudos  & 2021-01-14 & 2024-11-08 & 1025  & -18   \\ \hline
        tomo   & 2020-01-01 & 2023-12-01 & 1461  & -31   \\ \hline
        wtc    & 2020-01-01 & 2022-06-17 & 1461  & 168   \\ \hline
        zks    & 2021-01-07 & 2024-01-26 & 1089  & 25    \\ \hline
    \end{tabular}
\end{table}

\begin{table}[h]
    \centering
    \caption{Available data points using optimal date splitting setup 0614}
    \label{tab:av_data_optimal_split}
    \begin{tabular}{|l||l|l|r|r|l|}  \hline
          &  &  &  &  & Optimal \\
          &  &  & Before & After & Split \\
          & First Seen & Last Seen & Split & Split & Date \\ \hline  \hline
        aion   & 2019-06-22 & 2023-03-04 & 1215 & 135 & 2022-10-19 \\ \hline
        ant    & 2019-06-22 & 2024-11-08 & 1768 & 196 & 2024-04-25 \\ \hline
        crpt   & 2019-06-22 & 2024-09-19 & 1723 & 191 & 2024-03-11 \\ \hline
        cudos  & 2021-01-14 & 2024-11-08 & 1252 & 139 & 2024-06-21 \\ \hline
        tomo   & 2019-06-22 & 2023-12-01 & 1459 & 162 & 2023-06-21 \\ \hline
        wtc    & 2019-06-22 & 2022-06-17 & 1638 & 182 & 2023-12-17 \\ \hline
        zks    & 2021-01-07 & 2024-01-26 & 1000 & 111 & 2023-10-06 \\ \hline
    \end{tabular}
\end{table}

\subsection{On Train and Test Data}
In Table~\ref{tab:traintest}, we present the results for the quality of our learned models. We cannot detect any big impact by changing the fixed-date split to the relative split to the best of our knowledge. Even if we extend the width to a little more data, we keep the tp-tn-rate larger than $0.9$. For data with less than 90 days of data, we can't predict so we just skip the prediction for those.
\begin{table}[h]
    \centering
    \caption{Test results on training data}
    \begin{tabular}{|c||c|c|c|c|c|c|}  \hline
        Setup Id & roc auc & fp & fn & tp & tn & skip \\ \hline \hline
        0602 & 0.9967 & 96 & 3 & 18 & 0 & 4 \\ \hline
        0603 & 0.9876 & 46 & 0 & 21 & 50 & 4 \\ \hline
        0604 & 0.9928 & 96 & 3 & 18 & 0 & 4 \\ \hline
        0605 & 0.9785 & 38 & 0 & 21 & 58 & 4 \\ \hline
        0610 & 0.9268 & 41 & 3 & 35 & 0 & 3 \\ \hline
        0614 & 0.9268 & 41 & 4 & 34 & 0 & 3 \\ \hline
    \end{tabular}
    \label{tab:traintest}
\end{table}

\subsection{Unseen Data}
\begin{figure}[h]
    \centering        
    \includegraphics[width=0.45\textwidth]{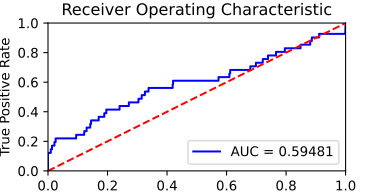}
    \caption{ROC for setup 0614}
  \label{fig:roc_auc_full}
\end{figure}

As we have presented so far, we have very good ROC AUC values by predicting the well-known train and test data. The most interesting step is when we apply our model also to unseen data. In Table~\ref{tab:unseen} we present the results by predicting all our available crypto assets data. False positives (fp) describe that an alive coin was predicted correctly; true positives (tp) counts a dead coin was labelled correctly. The ROC AUC \qeng{n.d.} cannot be calculated, since division by zero is not defined. In Figure~\ref{fig:roc_auc_full} we can see that we are a small bit better than randomly flipping a coin to predict. The model predicts in a few cases that coins are correct, even the 41 we used for training and testing. In contrast to that fact, we also have 78 percent true negative predictions in our results.
\begin{table}[h]
    \centering
    \caption{Results for all coins}
    \begin{tabular}{|c||c|c|c|c|c|c|}  \hline 
        Setup Id & roc auc & fp & fn & tp & tn & skip \\ \hline \hline
        0602 & 0.6454 & 210 & 2 & 17 & 486 & 52 \\ \hline
        0603 & 0.5646 & 54 & 0 & 19 & 642 & 52 \\ \hline
        0604 & 0.6124 & 190 & 2 & 17 & 506 & 52 \\ \hline
        0605 & n.d. & 98 & 21 & 0 & 607 & 44 \\ \hline
        0610 & 0.6015 & 121 & 3 & 35 & 618 & 4 \\ \hline
        0614 & 0.5948 & 126 & 4 & 34 & 613 & 4 \\ \hline
    \end{tabular}
    \label{tab:unseen}
\end{table}

\begin{figure}[h]
    \centering        
    \includegraphics[width=0.57\textwidth]{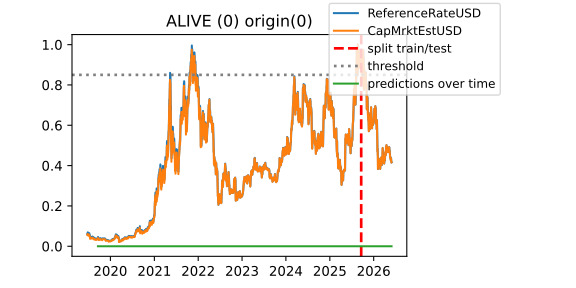}
    \caption{Historical data for eth}
  \label{fig:eth_hist_series}
\end{figure}

\begin{figure}[h]
    \centering        
    \includegraphics[width=0.57\textwidth]{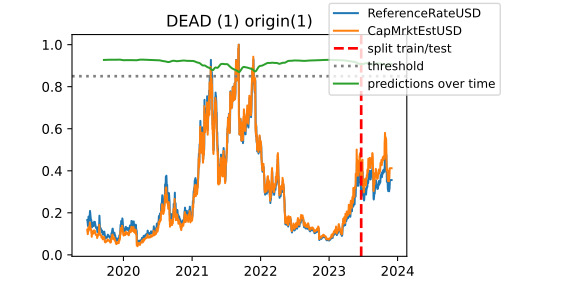}
    \caption{Historical data for tomo}
  \label{fig:tomo_hist_series}
\end{figure}

Finally, we analysed the given time series depending on the prediction and the correctness of the prediction. Figures~\ref{fig:eth_hist_series} and~\ref{fig:tomo_hist_series} show, for example, historical data for false-positive predicted for crypto asset eth and true-positive predicted for crypto asset tomo. On the x-axis we see time indexes (daily) in the y-axis, we present the normalised metrics CapMrktEstUSD and ReferenceRateUSD we have used for training and test our LSTM model. On top of the figures, we show the prediction dead or alive and what the original label was before training, testing, and validation. Except for asset ohm and ignoring the time information, we have no intention to explain with the time series data why this is a dead asset or why not, and also it does not explain the wrong prediction. That´s why we did another additional analysis with our data.

\subsection{On Series of a crypto asset}
\begin{figure}[h]
    \centering        
    \includegraphics[width=0.57\textwidth]{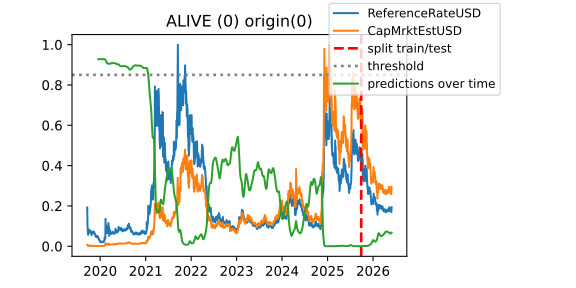}
    \caption{Historical data for hbar}
  \label{fig:hbar_hist_series}
\end{figure}

To get an idea why our model predicts in some cases wrong, we also predicted for all slicing windows of the assets, and added these prediction to the time series data of the assets. Figure~\ref {fig:hbar_hist_series} shows, on the one hand, a false positive history data of hbar and additionally the prediction for the full time series and the selected threshold. In Figure~\ref{fig:tomo_hist_series} an example for true positive detected asset tomo. For asset eth (Figure~\ref{fig:eth_hist_series}) we got an almost stable curve close to zero, and in contrast, for asset tomo, absolutely over the threshold. The most interesting finding was for the assets like hbar. The curve is changing its values over time. This green \qeng{life-line} was a surprising result, and we wonder how we can use this for further predictions.

%% file: 04_future_work_conclusion.tex
\section{Future Work and Conclusion}
Several avenues for future work emerge from the findings and limitations of this work.

Feature expansion represents the most immediate opportunity. Incorporating daily trading volume, which is the most commonly cited indicator of market health in the dead coin literature, would likely improve discriminative performance. Beyond volume, on-chain metrics such as active address counts and transaction values, where available, could provide direct evidence of network utilization. Off-chain signals, including the frequency of GitHub commits and social media sentiment, have been shown to correlate with project viability in related studies and could be integrated for assets where such data exist. A practical approach would be to train models at multiple tiers of feature availability: a base model using only price and market capitalisation, an intermediate model adding volume, and an enriched model incorporating on-chain and off-chain features for assets where these are accessible. Horizon-specific labeling is a second important direction. The current study assigns a static coin-level label to all windows, but a more informative formulation would be to predict whether an asset will become inactive within a specific future horizon, for example, within 30, 60, or 90 days. Also we have to consider, that data labeled with \qeng{dead} may not have hidden signals in the first parts of the time series. This time-to-event framework, which can be implemented either as a classification problem with horizon-specific labels or as a survival analysis problem using Cox proportional hazard models or neural network-based survival models, would provide actionable temporal information and reduce the label noise associated with coin-level labeling.
The within-study ablation confirmed that the discriminative signal is largely attributable to feature separability rather than temporal architecture alone. Future ablation studies should extend this comparison to include GRU and BiLSTM architectures, which were excluded on theoretical grounds, in order to empirically quantify the recall advantage of standard LSTM over these sequential alternatives under the fixed-window deployment setting. Additionally, expanding the dataset beyond 41 dead coins, by relaxing the market capitalization filter or integrating data from multiple providers, would improve generalization across the full spectrum of failure modes. Model interpretability should be enhanced through the application of post-hoc explanation methods. Techniques such as SHAP (Shapley Additive Explanations), integrated gradients, or attention-weight visualization could identify which time steps and features within a 90-day window contribute most to the death probability. Such analyses would complement the qualitative insights from the pre-death signature analysis and provide more actionable information for end users. The work of Badar et al.~\cite{badar2025}, which demonstrated the application of SHAP to a hybrid CNN-LSTM model for price prediction, provides a methodological template for this extension.

Operational deployment and lifecycle modeling also warrant further investigation. A production system would need to address model retraining frequency, threshold recalibration, and alert aggregation. Persistence rules, where an asset is flagged only after exceeding the threshold for consecutive days, could reduce false positives. In the longer term, as data archival practices improve and richer project-level data becomes available, revisiting the multi-stage lifecycle concept becomes a natural extension, with the binary classifier developed here serving as the foundation for a hierarchical approach.

The study makes several contributions to the emerging literature on cryptocurrency failure prediction. It demonstrates that a minimal feature set of only two market variables, combined with a sequence learning architecture, can achieve discriminative performance that equals or exceeds results reported in prior studies that used richer feature sets or more complex model ensembles. It provides empirical evidence that pre-death market behavior is characterized by gradual value erosion and elevated volatility rather than sudden catastrophic collapse, supporting a temporal pattern that recurrent neural networks are well positioned to capture. It documents and justifies the transition from a conceptually appealing but practically infeasible multi-stage lifecycle model to a data-driven binary classification approach, contributing to the methodological discourse on how to model cryptocurrency project outcomes given current data constraints.

However, it is important to acknowledge the limitations of this work. The limited number of dead coins in the sample, the single-provider data source, the coin-level labeling strategy, and the restricted two-feature input set constrain the strength of the conclusions. It is recommended that an additional experiment and analysis be conducted against the logistic regression, random forest and XGBoost baselines. These findings define a clear agenda for future research: expanded datasets, horizon-specific labels, multi-provider validation, richer feature sets, and rigorous ablation studies in settings where temporal modelling's incremental benefit can be more precisely isolated. In sum, this work provides evidence that the market traces of cryptocurrency failure are detectable well before an asset becomes fully illiquid, and that LSTM networks offer an effective means of capturing these traces from minimal input data. As the cryptocurrency ecosystem continues to grow and evolve, the ability to identify at-risk assets in a timely and scalable manner will become increasingly important for investors, exchanges, and regulators. The framework developed here offers a foundation upon which more sophisticated, feature-rich, and operationally deployed early warning systems can be built.